\documentclass[osajnl,twocolumn,
preprintnumbers,elsart]{revtex4-1}
\usepackage{amsmath}
\usepackage{cases}
\usepackage{mathrsfs}
\usepackage{graphicx}
\usepackage{dcolumn}
\usepackage{bm}
\usepackage[center]{subfigure}
\usepackage{color}
\usepackage{float}

\begin{document}

\title{Tunable storage of optical pulses in a tailored Bragg-grating
structure}
\author{Shenhe Fu$^{1,2}$, Yongyao Li$^{3}$, Yikun Liu$^{1}$}
\email{ykliu714@gmail.com }
\author{Jianying Zhou$^{1}$}
\author{Boris A. Malomed$^{2}$}
\address{
$^{1}$State Key Laboratory of Optoelectronic Materials and Technologies, Sun Yat-sen University, Guangzhou 510275, China \\
$^{2}$Department of Physical Electronics, School of Electrical Engineering,Faculty of Engineering, Tel Aviv University, Tel Aviv 69978, Israel\\
$^{3}$Department of Applied Physics, South China Agriculture University, Guangzhou 510642, China}

\begin{abstract}
Scenarios for controllable creation, trapping and holding of single and
multiple solitons in a specially designed nonlinear Bragg grating (BG) are
proposed. The setting includes a chirped BG segment, which is linked via a
local defect to a uniform BG with a built-in array of defects. A parabolic
relation between the trapping position of the incident soliton and its power
is obtained. Simultaneous trapping of two and three solitons at different
locations is demonstrated too.
\end{abstract}

\maketitle

\noindent
OCIS (060. 3735) Fiber Bragg gratings; (060.5530) Pulse propagation and temporal solitons;
(060.1810) Buffers, couplers, routers, switches, and multiplexers

\section{Introduction}

A possibility to dramatically reduce the velocity of ultrashort light pulses
in various optical materials \cite{Safavi2011, Bigelow2003, Li2012} has
drawn a great deal of attention, since the observation of slow \cite{Hau1999}
and completely halted light \cite{Phillips2001} in experiments. However, the
controlled storage of this ultrashort light pulse has been a challenge,
which hinders their use in broadband optical signal processing, such as
ultrafast coding, decoding and multiplexing. Different settings were
proposed to realized the slow light, such as electromagnetically induced
transparency \cite{Safavi2011, Kasapi1995, Khurgin2005}, simulated Brillouin
\cite{Okawachi2005} and Raman \cite{Sharping2005} scattering, coherent
population oscillations \cite{Bigelow2003}, dispersion engineering in
photonic-crystal waveguides \cite{Li2012, Vlasov2005}, and others. Among
these techniques, Bragg-grating (BG) structures, which induce very strong
dispersion in a vicinity of the photonic bandgap \cite{Mok2006}, are used
too, for the generation the slow light due to its tunability and
cascadability in practical applications.

The relation between the delay and bandwidth of pulses was previously
experimentally demonstrated in a linear resonant systems \cite{Lenz2001}. To
support the slow propagation of non-spreading pulses in the BG, the Kerr
nonlinearity may be used, as it is not subject to the limitation of the
trade-off between the pulse delay and bandwidth, which is inherent in the
linear settings \cite{Lenz2001}. Standing and moving robust light pulses (BG
solitons), whose velocity may be orders of magnitude smaller than the
velocity of light in the host material, were studied in detail theoretically
\cite{Christodoulides1989, Wabnitz,Sipe,Kozhekin1998, Li2006, Xiao2003,
Mak2004}. Producing quiescent pulses by collisions of counterpropagating BG\
solitons \cite{Mak2003} was predicted too.

In experiments, BG solitons, with the velocity corresponding to $76\%$ of
the speed of light in the uniform medium, were first produced in Ref. \cite%
{Eggleton1996}. Subsequently, a number of experiments \cite{Mok2006,
Longhi2005, Janner2005, Qian2009} were carried out to investigate slow-light
effects in BG structures. In particular, high intensity of the pulse leads
to a change of the local refractive index via the Kerr nonlinearity, which
helps to enhance the retardation of pulses propagating in the BG structure.
Recently, large femtosecond Kerr nonlinearity was experimentally
demonstrated in cholesteric liquid crystals \cite{Song2013}. Its large Kerr coefficient would
considerably reduce the high intensity. The lowest experimentally
demonstrated velocity of solitons in BGs is $0.16c_{0}$ \cite{Mok2006},
where $c_{0}$ is the speed of light in vacuum.

However, there remain challenging problems impeding the use of the slow
light pulses in BGs. First, the creation of very slow or standing optical
modes in BG structures has not yet been reported in experiments. Second, to
the best of our knowledge, no work has demonstrated, as yet, controllable
trapping and storage of single and multiple pulses at different positions in
BGs, a reason being that it is difficult to create suitable input pulses for
this purpose, using uniform BGs. Indeed, as the optical fields inside the BG
include forward- and backward-traveling waves, generating such very slow or
quiescent pulses requires to couple forward and backward waves with nearly
equal powers into the BG waveguide, doing which for the backward wave being
obviously difficult.

Recently, we have proposed a specially tailored structure, built of a
linearly chirped BG segment linked, through a local defect, to a uniform
grating \cite{Fu2013}. This setting demonstrates a possibility to create
very slow or standing stable pulses, the advantage being that the pulse
coupled into the uniform grating may be manipulated by means of the chirped
segment. This scheme makes it possible to prepare the right mix of forward-
and backward-traveling fields. Using it, the creation of BG solitons with
extremely small and zero velocities was predicted, offering an essential
improvement of previously published results \cite{Qian2009, Shnaiderman2011}.

In this work, we aim to develop the above-mentioned setting by introducing a
periodic set of defects into the uniform grating, to achieve controllable
trapping and storage of stable single and multiple pulses at different
positions in the BG. In section II, we present the necessary model, which is
based on nonlinear coupled-mode equations (CMEs) for the BG. In section III,
the operation of the system is demonstrated by means of CME simulations. The
paper is concluded by Section IV.

\section{The model}

The purpose of introducing a chirped BG before the uniform grating with a
periodic array of defects is to provide a suitable setting for the
generation of slow pulses, which can be subsequently trapped in the main
section of the system. First, the setting without the defect lattice is
defined by means of the spatially modulated profile of the local refractive
index \cite{Fu2013}: $n_{0}[1+2\Delta n\cos(2\pi z(1+Cz)/\Lambda_{0})]$ for (%
$0\leq z<L_{c}$), and $n_{0}[1+2\Delta n\cos(2\pi z(1+CL_{c})/\Lambda_{0})]$
for ($L_{c}\leq z\leq L$). Here $z$ is the propagation distance, $L$ the
total length of the setting, $L_{c}$ the length of the chirped Bragg
gratings, and $C$ the chirp coefficient. Further, $\Lambda _{0}$ is the BG
period at the input edge, $n_{0}$ the average refractive index, and $\Delta
n $ the amplitude of the refractive-index modulation.

The nonlinear CME system is an accurate model for the propagation of optical
pulses in BGs with the Kerr nonlinearity and inhomogeneity of the refractive
index \cite{Sipe,Huang,Broderick,Irina}:
\begin{gather}
\pm i\frac{\partial E_{f,b}}{\partial z}+\frac{i}{v_{g}}\frac{\partial
E_{f,b}}{\partial t}+\delta (z)E_{f,b}+\kappa (z)E_{b,f}  \notag \\
+\gamma (|E_{f,b}|^{2}+2|E_{b,f}|^{2})E_{f,b}=0,  \label{coupled}
\end{gather}%
where $t$ is time, $E_{f}$ and $E_{b}$ are amplitudes of the
forward-traveling and backward-traveling waves, respectively, $%
v_{g}=c_{0}/n_{0}$ is the group velocity in the material of which the BG is
fabricated, and $\gamma =n_{2}\omega /c$ is the nonlinearity coefficient,
with $\omega $ being the carrier frequency and $n_{2}$ the Kerr coefficient.
The strength of the coupling between forward and backward waves is $\kappa
=\pi \Delta n/\Lambda _{0}$, while $\delta (z)$, which represents the local
wavenumber detuning, is
\begin{numcases}{\delta(z)=}
\delta_0-2\pi Cz/\Lambda_{0}, & for $0\leq z<L_{c}$,  \notag \\
\delta_0-2\pi CL_{c}/\Lambda_{0}, & for $L_{c}\leq z\leq L$, \label{deltafun}
\end{numcases}where $\delta _{0}\equiv 2\pi n_{0}/\lambda -\pi /\Lambda _{0}$
is the detuning at the left end of chirped segment. The nonlinear terms in
Eqs. (\ref{coupled}) account for the self- and cross-phase modulation,
respectively. We denote the chirped segment of length $L_{c}$ as chirped-BG
(the chirped Bragg grating), and the uniform one as uniform-BG (the uniform
Bragg grating), the length of which is $L-L_{c}$. The setting implies that
the local BG period decreases with $z$ in the chirped-BG, hence the nominal
Bragg wavelength is gradually shifted to smaller values for pulses running
through the chirped grating. In other words, the wavelength of the input
pulse, originally taken near the blue edge of the bandgap, drifts into the
depth of the bandgap in the course of the propagation of the pulse \cite%
{Fu2013}. As a consequence, the power of the forward-propagating wave is
gradually converted into the backward wave, helping to introduce the balance
between the powers of the two waves, which is necessary for the creation of
a very slow BG soliton in the uniform-BG \cite%
{Christodoulides1989,Wabnitz,Sipe}. The soliton's velocity is determined by
the remaining imbalance between the powers:
\begin{equation}
V_{sol}=v_{g}(1-f)/(1+f)  \label{f}
\end{equation}%
\begin{equation*}
f(t)=\frac{\int_{0}^{L}|E_{b}(z,t)|^{2}dz}{\int_{0}^{L}|E_{f}(z,t)|^{2}dz}.
\end{equation*}%

\begin{figure}[t]
\centering\includegraphics[width=8cm]{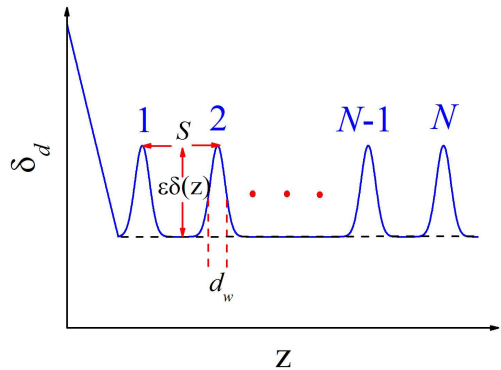}
\caption{(Color online) A schematic of the system, built of the linearly
chirped BG segment (on the left-hand side) followed by the uniform grating
with an inserted periodic array of local defects, which is described by the
set of parameters $(S,d_{w},\protect\varepsilon )$.}
\label{fig1}
\end{figure}

By introducing a defect at the junction between the chirped-BG and
uniform-BG segments, one can create a standing soliton stably trapped by the
defect \cite{Fu2013}. Here, we introduce an array of defects in the
uniform-BG, as multiple positions at which the stable trapping should be
possible. For this purpose, we define the local detuning in the uniform-BG
region as
\begin{equation}
\delta _{d}(z)=\delta (z)\times \left[ 1+\varepsilon \sum_{n=1}^{N}\mathrm{%
exp}\left\{ -\frac{[z-(1+(n-1)S)]^{2}}{d_{w}^{2}}\right\} \right]
\label{delta}
\end{equation}%
with $\varepsilon $ being the depth of each defect, $N$ the total number of
defects, $S$ the spacing between them, and $d_{w}$ the width of defect. The
array is displayed in Fig. \ref{fig1}, which shows $\delta _{d}(z)$ as a
function of $z$. Thus, the defect array is determined by a parameter set, $%
(S,d_{w},\varepsilon )$. The respective shape of the effective potential for
solitons, $V(z)$, which is opposite to $\delta _{d}(z)$ \cite%
{Malomed2003,Peter}, predicts that, when the pulse's kinetic energy, $K$, is
smaller than the local height of the effective potential, $V_{0}$, the pulse
cannot pass the local potential maximum. If $K$ slightly exceeds $V_{0}$, it
is expected that the pulse may pass the maximum into the adjacent potential
well at a small velocity, and get trapped there \cite{Peter}. Below, this
possibility is confirmed by systematic simulations of Eqs. (\ref{coupled}).

\section{Numerical analysis}

\subsection{Controllable trapping of the single pulse}

In this section, systematic simulations of the model are reported for
physical parameters of silicon, whose Kerr coefficient and average
refractive index are $4.5\times 10^{-14}$ cm$^{2}/$W and $n_{0}=3.42$ \cite%
{Dinu2003}. The amplitude of the refractive index-modulation is taken as $%
\Delta n=0.006$, a typical chirp coefficient in the chirped-BG segments $%
C=2.5064\times 10^{-4}$ cm$^{-1}$, and the BG period at the input edge of
the sample is fixed to be $\Lambda _{0}=154.1$ nm. Simulations based on Eqs.
(\ref{coupled}) and (\ref{delta}) were carried out with the boundary
condition corresponding to the Gaussian pulse,
\begin{equation}
E_{f}(z=0,t)=E_{0}\exp \left[ -(t/t_{0})^{2}\right] ,  \label{in}
\end{equation}%
with temporal width $t_{0}=16$ ps, launched into the system at carrier
wavelength $1053$ nm, which is placed near the blue edge of the photonic
bandgap. Further, the total length of the system and the length of the
chirped-BG segment are fixed as $L=1.54$ cm and $L_{c}=0.193$ cm,
respectively (below, the influence of $L_{c}$ on the pulse storage is
discussed), thus the length of the uniform-BG segment is $1.347$ cm. The
number of defects in the array is $N=10 $, with individual defect
characterized by parameters $(S,d_{w},\varepsilon )=(0.132$ cm, $50$ $%
\mathrm{\mu }$m, $0.04)$, i.e., the space between the defects, $S$, is much
larger than the width of each one, $d_{w}$. In comparison with the spatial
width of the pulse (the temporal duration of $16$ ps corresponds to $4.8$ mm
in space) the defect of width $50$ $\mathrm{\mu }$m may indeed be considered
as a point-like object.

\begin{figure}[t]
\centering
\subfigure[]{\includegraphics[width=4cm, height=3cm]{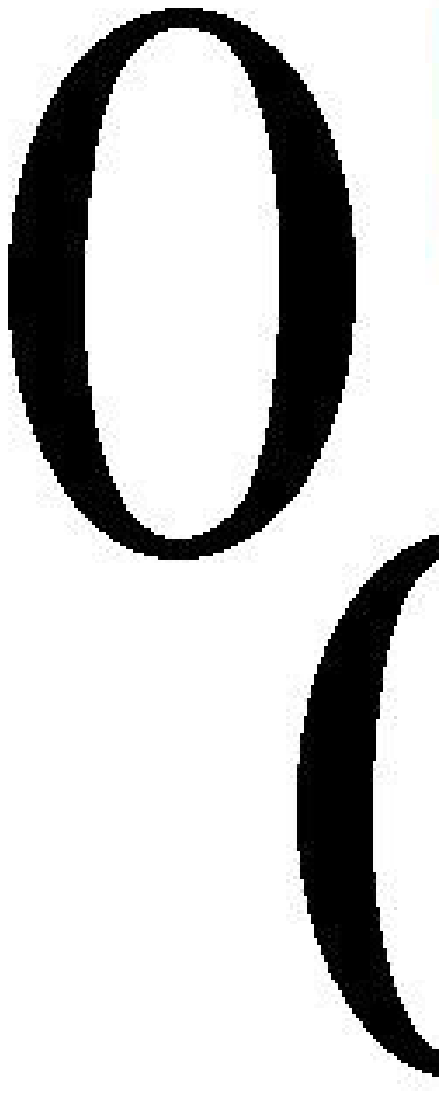}} %
\subfigure[]{\includegraphics[width=4cm, height=3cm]{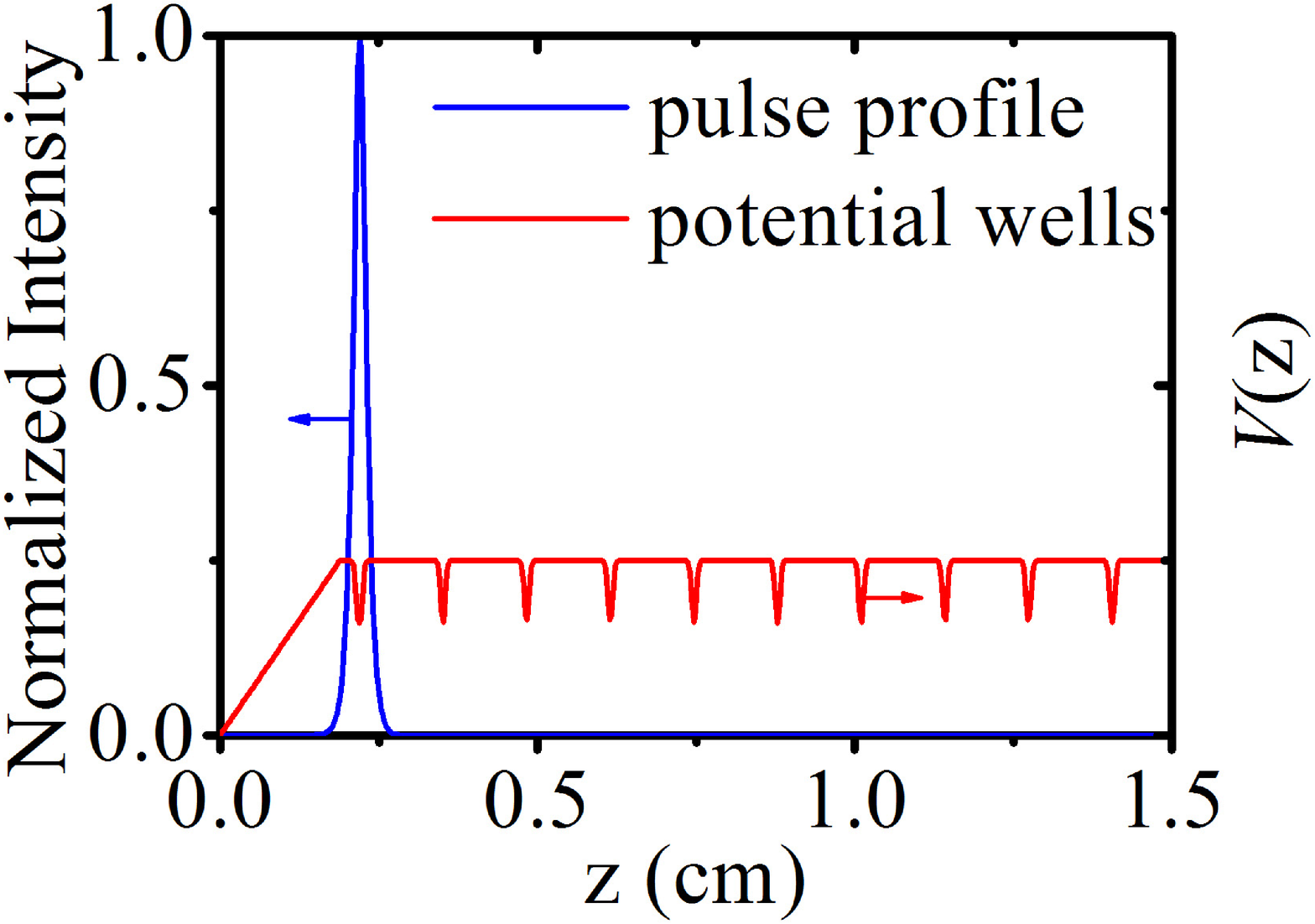}} %
\subfigure[]{\includegraphics[width=4cm, height=3cm]{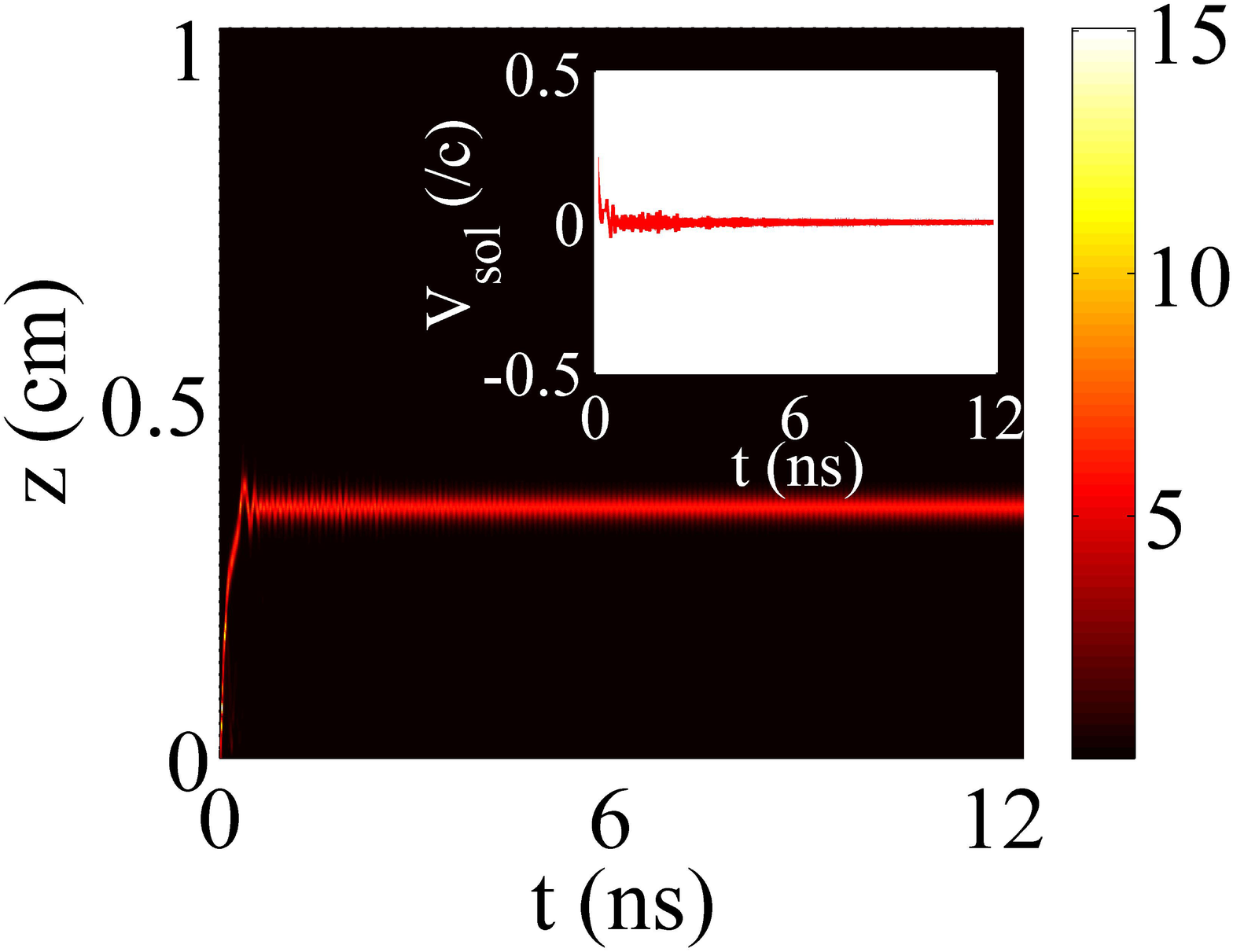}} %
\subfigure[]{\includegraphics[width=4cm, height=3cm]{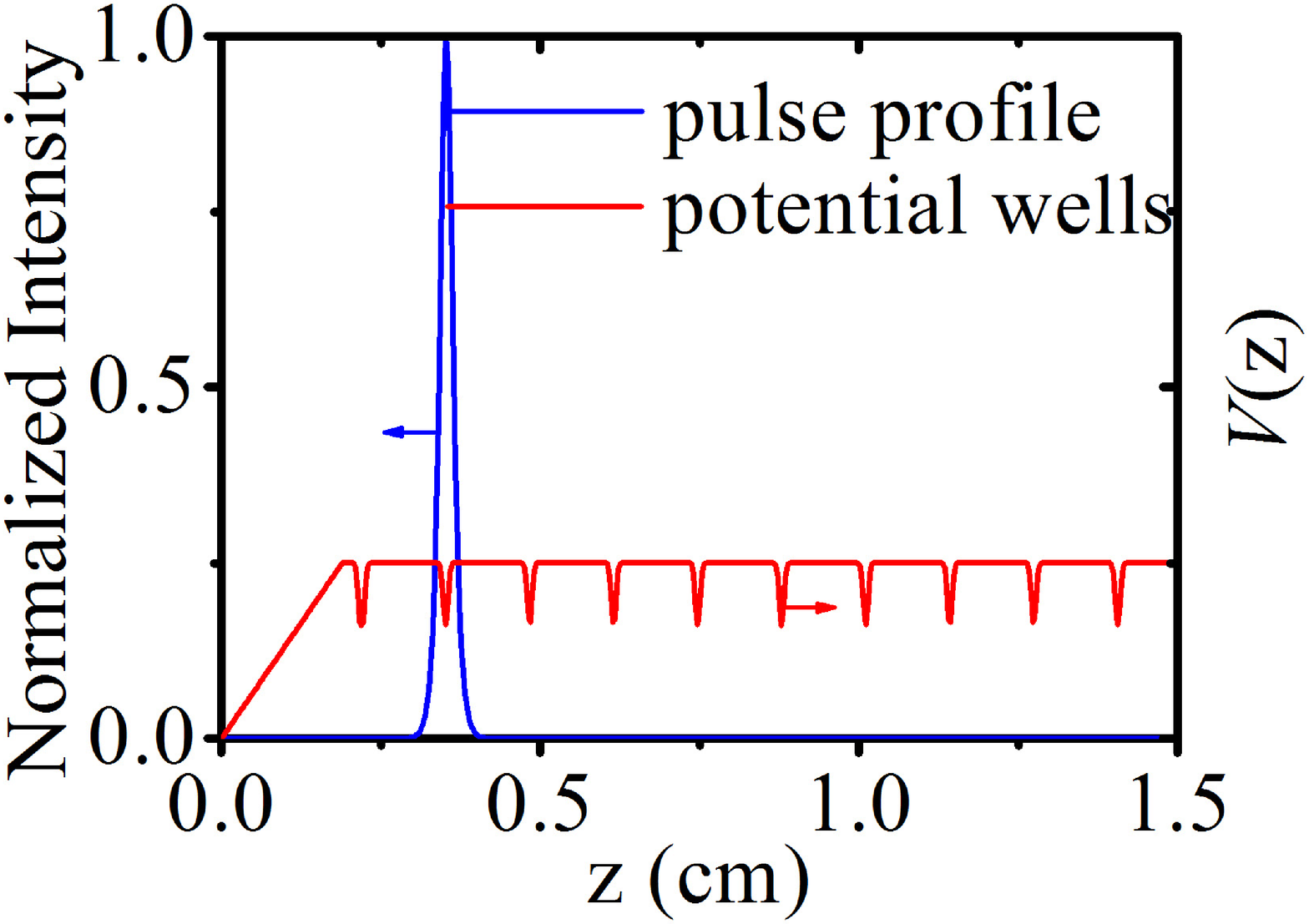}} %
\subfigure[]{\includegraphics[width=4cm, height=3cm]{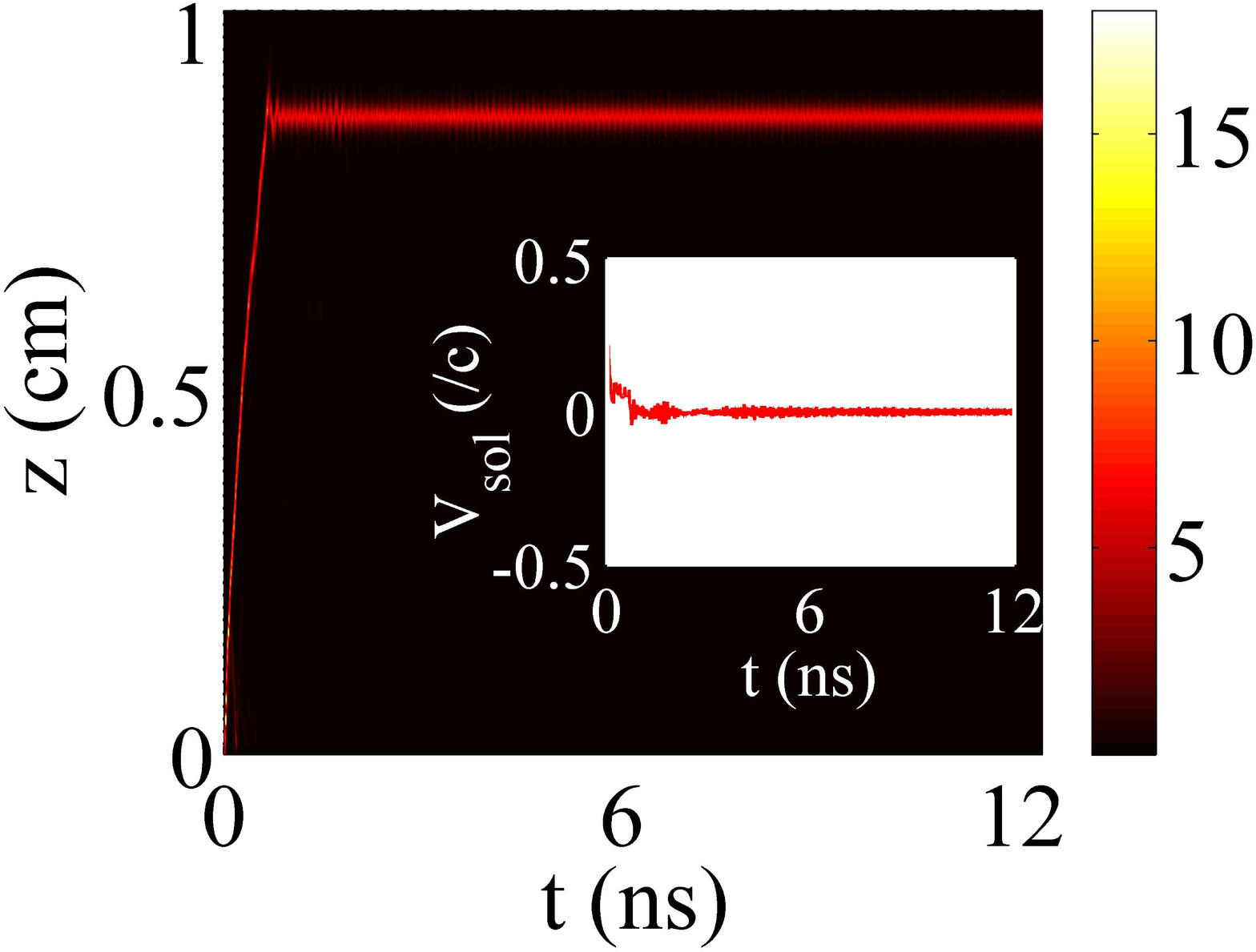}} %
\subfigure[]{\includegraphics[width=4cm, height=3cm]{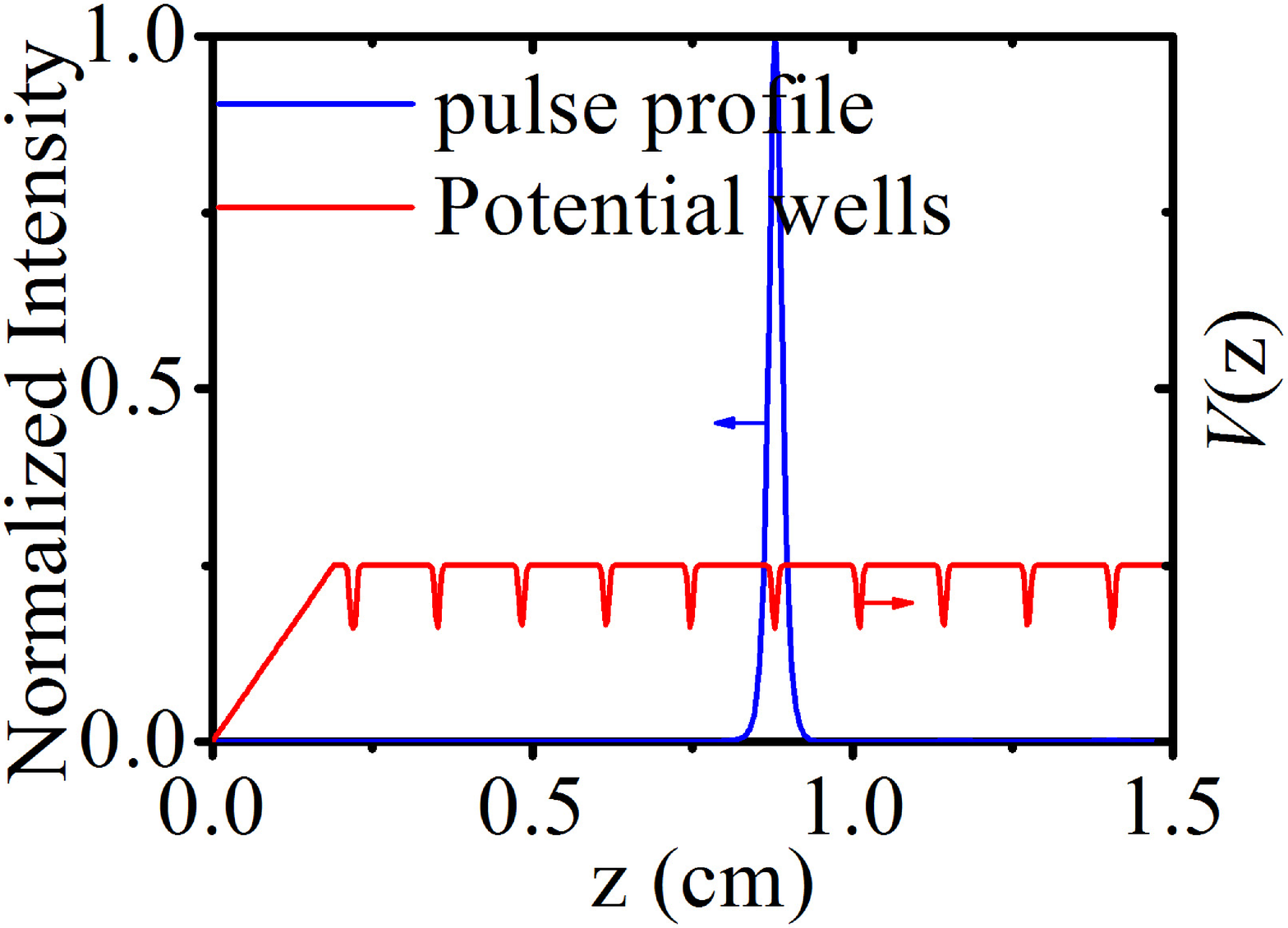}}
\caption{(Color online) The nonlinear propagation of pulses in the system
with the defect array with ($S,d_{w},\protect\varepsilon )=(0.132~$cm$,50$ $%
\mathrm{\protect\mu }$m, $0.04)$. The first and second columns demonstrate,
severally, the intensity evolutions and the corresponding profiles at $t=12$
ns, for three different injection intensities: (a,b) $2.07$ GW$/$cm$^{2}$,
(c,d) $2.72$ GW$/$cm$^{2}$, (e,f) $3.36$ GW$/$cm$^{2}$. Insets of (a), (c),
(e) represent the evolutions of the pulse's velocity $V_{sol}$, calculated by
Eq. (\protect\ref{f}).}
\label{fig2}
\end{figure}

Simulations of Eqs. (\ref{coupled}) were performed, by means of the
fourth-order Runge-Kutta method. The intensities $I_{P}$ used in simulations are below
the damage threshold \cite{Pronko1998} [$I_{P}$ is the peak intensity of pulse,
defined as $I_{P}=|E_{0}|^{2}$, see Eq. (\ref{in})].  First, setting the initial pulse intensity $%
I_{P}$ to $2.07$ GW$/$cm$^{2}$, trapping of a stable
pulse by the first defect (i.e., the formation of a standing optical
soliton) is observed, as shown in Figs. \ref{fig2}(a,b). These results
demonstrate that the Kerr nonlinearity in the BG, at the present (quite
realistic) intensity level not only compensates the pulse's dispersion in
the vicinity of the edge of photonic bandgap, securing the formation of the
gap soliton, but also shifts the photonic bandgap to a longer wavelength,
which enables the pulse to pass the chirped-BG segment, before getting
trapped in the uniform-BG section. Indeed, when the pulse reaches the
junction between chirped-BG and uniform-BG, the power of the forward and
backward components approximately equal, giving rise to a nearly halted
pulse, as seen from Eq. (\ref{f}), which may be readily captured by the
first potential well.

The evolution of the pulse's overall velocity, $V_{sol}$, defined according to
Eq. (\ref{f}), is additionally shown in the inset of Fig. \ref{fig2}(a). It
is seen that $V_{sol}$ rapidly drops at $t<3$ ns, due to the transfer from
forward component to the backward one in the course of passing the
chirped-BG segment. After that, $V_{sol}$ approaches zero, suggesting the
trapped of the soliton in the potential well.

With the incident-pulse intensity increased to $I_{P}=2.72$ GW$/$cm$^{2}$,
trapping of a stable soliton in the second potential well is observed in
Figs. \ref{fig2}(c,d). These results reveal that the pulse with this level
of the intensity passes the first defect, and then keep moving at a small
velocity, eventually being trapped by the second well. The further increase
of the intensity to $I_{P}=3.36$ GW$/$cm$^{2}$, as shown in Figs. \ref{fig2}%
(e,f), leads to the trapping of a stable soliton in the sixth potential
well.

\begin{figure}[t]
\centering
\subfigure[]{\includegraphics[width=4cm, height=3cm]{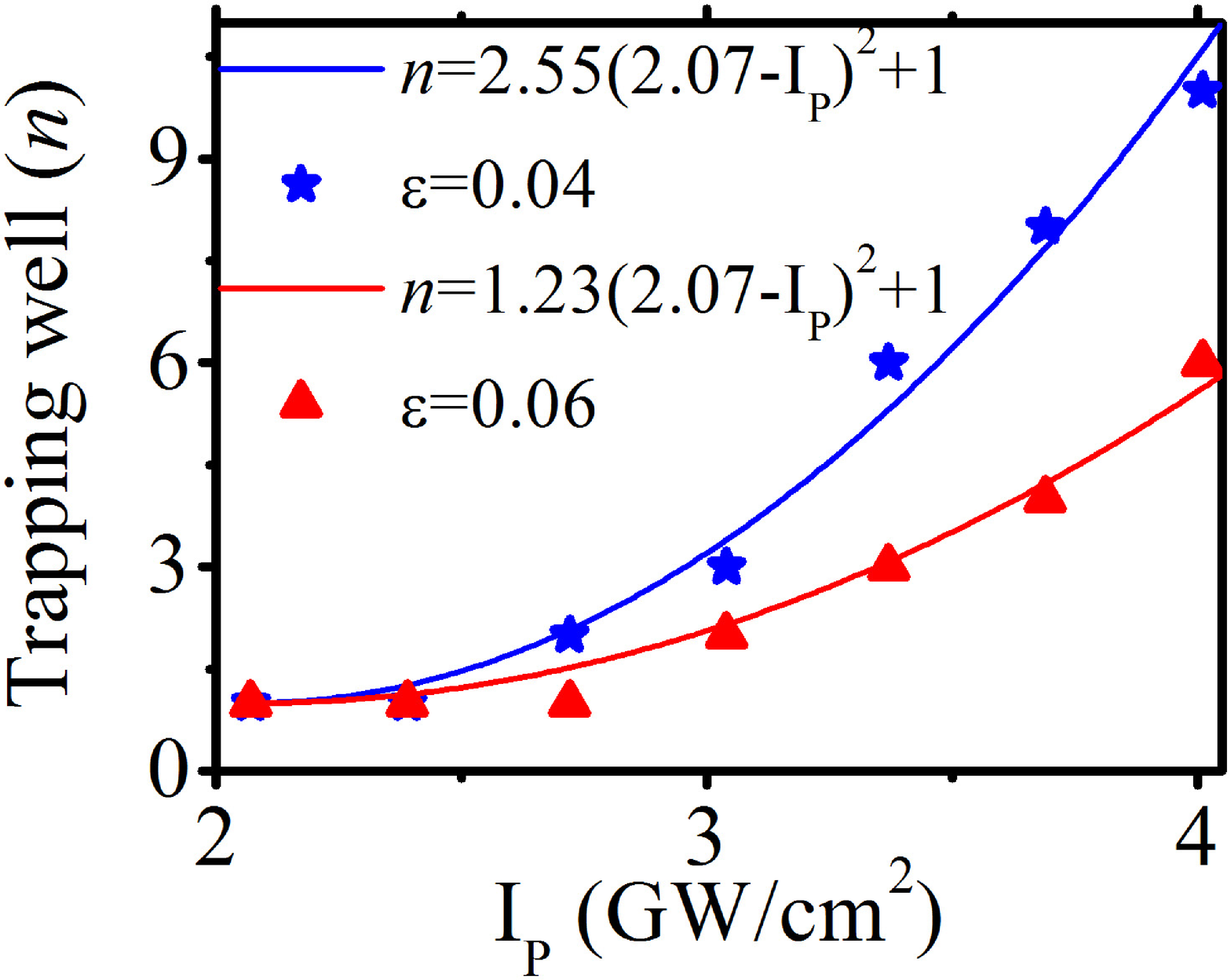}} %
\subfigure[]{\includegraphics[width=4cm, height=3cm]{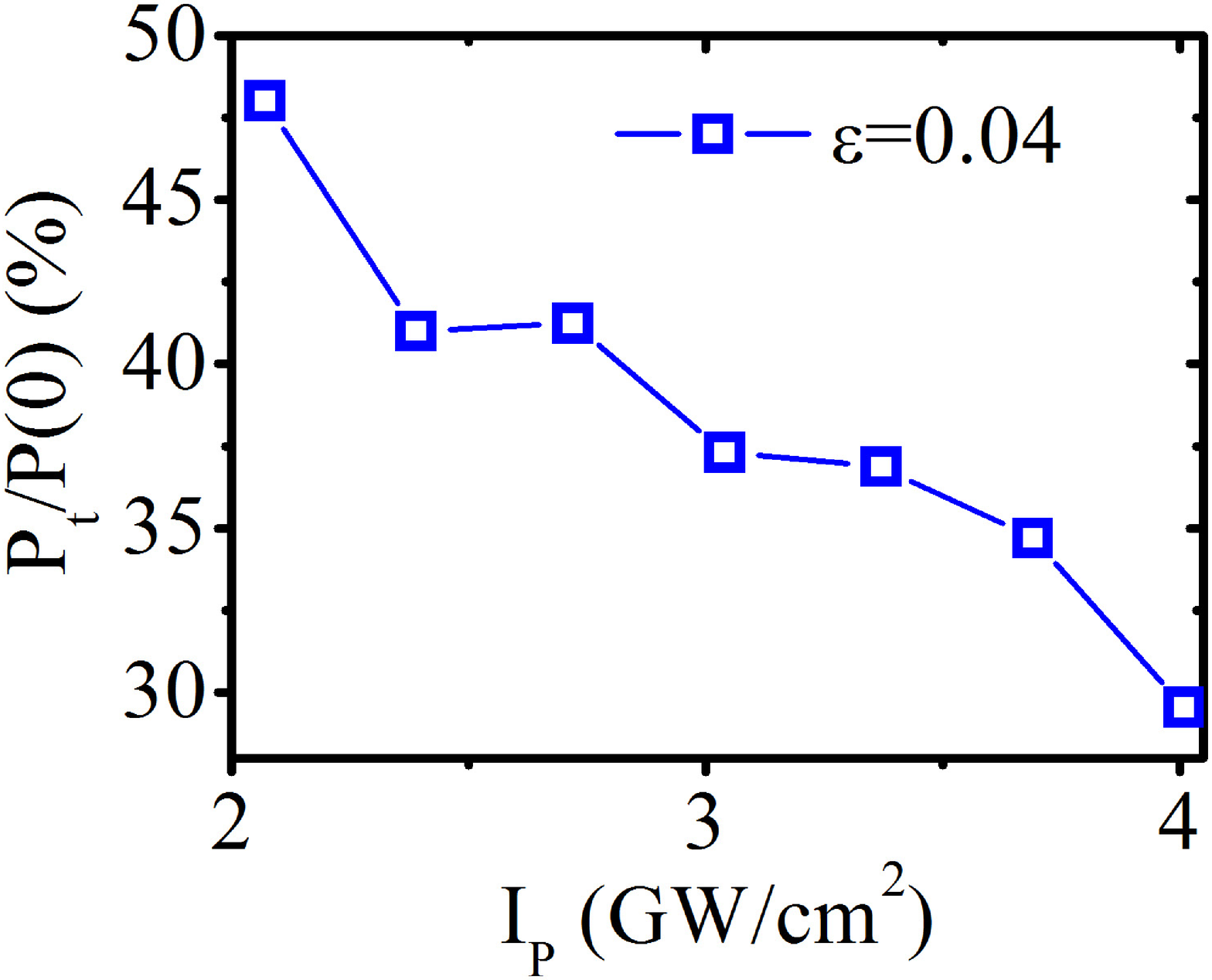}}
\caption{(Color online) (a) The relation between the trapping position and
the initial pulse's intensity $I_{P}$ for $(S,d_{w},\protect\varepsilon %
)=(0.132$ cm, $50$ $\mathrm{\protect\mu }$m, $0.04)$ and $(0.132$ cm$,50~%
\mathrm{\protect\mu }$m$,0.06)$ (blue and red curves, respectively). (b) The
ratio of the residual total power of the trapped pulse to the input power at
$\protect\varepsilon =0.04$, versus the initial intensity.}
\label{fig3}
\end{figure}

The above results suggest that the number of the trapping effective
potential well increases with the increase of the incident-pulse's intensity. To
clearly demonstrate this trend, simulations were carried out with different
intensities $I_{P}$, varying from $2$ GW$/$cm$^{2}$ to $4.05$ GW$/$cm$^{2}$,
for two different defect strengths, $\varepsilon =0.04$ and $\varepsilon
=0.06$. The outcome is depicted in Fig. \ref{fig3}(a), which shows the
trapping position, $n$, as a function of the intensity. In both cases, the
relation between $n$ and $I_{P}$ can be approximated by
\begin{equation}
n=1+a(I_{P}-I_{\mathrm{cr}})^{2},  \label{nI}
\end{equation}%
where $I_{\mathrm{cr}}$ is a critical value of the injected intensity, $I_{P}
$, for trapping the pulse in the first potential well, and $a$ is obtained
by fitting the numerical data. For instance, $a=2.55$ cm$^{4}/$(GW)$^{2}$
for $\varepsilon =0.04$, and $a=1.23$ cm$^{4}/$(GW)$^{2}$ for $\varepsilon
=0.06$. Although the critical intensities are practically identical in both
these cases, $I_{\mathrm{cr}}=2.07$ GW$/$cm$^{2}$, see Fig. \ref{fig3}(a),
there is a trend to decrease of $I_{\mathrm{cr}}$ with the decrease of $%
\varepsilon $. Indeed, in the limit of $\varepsilon \rightarrow 0$ any pulse
is able to pass the defect. Naturally, pulses stay trapped in a particular
potential well when $I_{P}$ takes values in some interval \cite{Fu2013}. For
example, at $\varepsilon =0.06$, the pulse is trapped by the first well for $%
2.07$ GW$/$cm$^{2}<I_{P}<2.72$ GW$/$cm$^{2}$.

The quadratic dependence displayed by Eq. (\ref{nI}) can be explained,
making use of the soliton's equation of motion. In the framework of the CME
system (\ref{coupled}), the soliton's momentum is defined as (see, e.g.,
Ref. \cite{Mak2004})
\begin{equation}
M=i\int_{-\infty }^{+\infty }\left( \frac{\partial E_{f}^{\ast }}{\partial z}%
E_{f}+\frac{\partial E_{b}^{\ast }}{\partial z}E_{b}\right) dz.  \label{M}
\end{equation}%
As follows from the boundary conditions (\ref{in}), the initial momentum of
the injected pulse scales as $M_{0}\sim I_{P}$. On the other hand, for a
relatively slow soliton, its velocity is proportional to the momentum, $%
V_{sol}\sim M$, see Eq. (\ref{f}) \cite{Mak2004}. Further, the radiative
braking force acting on the soliton, $F_{\mathrm{br}}$, can be estimated as
a loss of the momentum by the soliton passing a single defect, $\Delta M$,
times the number of defects passed in a unit of time, $\approx V_{sol}/S$, see
Eq. (\ref{delta}). The perturbation theory \cite{RMP} shows that $\Delta M$
is proportional to the time necessary for the soliton to pass the defect,
i.e., $\Delta M\sim d_{w}/V_{sol}$. Thus, we obtain $F_{\mathrm{br}}\sim
d_{w}/S$, which does not depend on velocity $V_{sol}$, and the corresponding
equation of motion, $dM/dt=-F_{\mathrm{br}}$, predicts that the soliton will
come to a halt, $M=0$, at time $t_{\mathrm{halt}}\sim M_{0}$. Finally, an
elementary mechanical analysis demonstrates that the total distance passed
by the soliton under the action of the constant braking force is $Z_{\mathrm{%
halt}}\sim M_{0}^{2}/F_{\mathrm{br}}\sim I_{P}^{2}$, which explains the
quadratic fit provided by Eq. (\ref{nI}).

The soliton relaxing to the eventually trapped stationary state loses a part
of its power through emission of linear waves (radiation). Accordingly, the
final value of the total power of the trapped optical pulse,
\begin{equation}
P_{\mathrm{t}}=\int_{\mathrm{trap}}\left[ \left\vert E_{f}(z,t\rightarrow
\infty )\right\vert ^{2}+\left\vert E_{b}(z,t\rightarrow \infty )\right\vert
^{2}\right] dz  \label{Pt}
\end{equation}%
(the integral is computed over a region where the soliton is eventually
trapped), normalized to the total input power, $P(0)$, is displayed, as a
function of the input intensity, $I_{P}$, in Fig. \ref{fig3}(b). A trend of $%
P_{\mathrm{t}}/P(0)$ to decrease with the increase of $I_{P}$ is explained
by the fact that the pulse with larger power passes a larger number of
defects, hence it is subject to stronger radiation losses. This curve also
shows a weak oscillatory behavior with the increase of $I_{P}$, ratio $P_{\mathrm{t}}/P(0)$
staying nearly constant in narrow intervals of $I_{P}$. This phenomenon can be
explained: pulses with $I_{P}$ taking values in some intervals, which indeed are
narrow, stay eventually trapped in the same potential well. Past these intervals,
the ratio $P_{\mathrm{t}}/P(0)$ again decreases with the increase of $I_{P}$.

\begin{figure}[t]
\centering
\subfigure[]{\includegraphics[width=4cm, height=3cm]{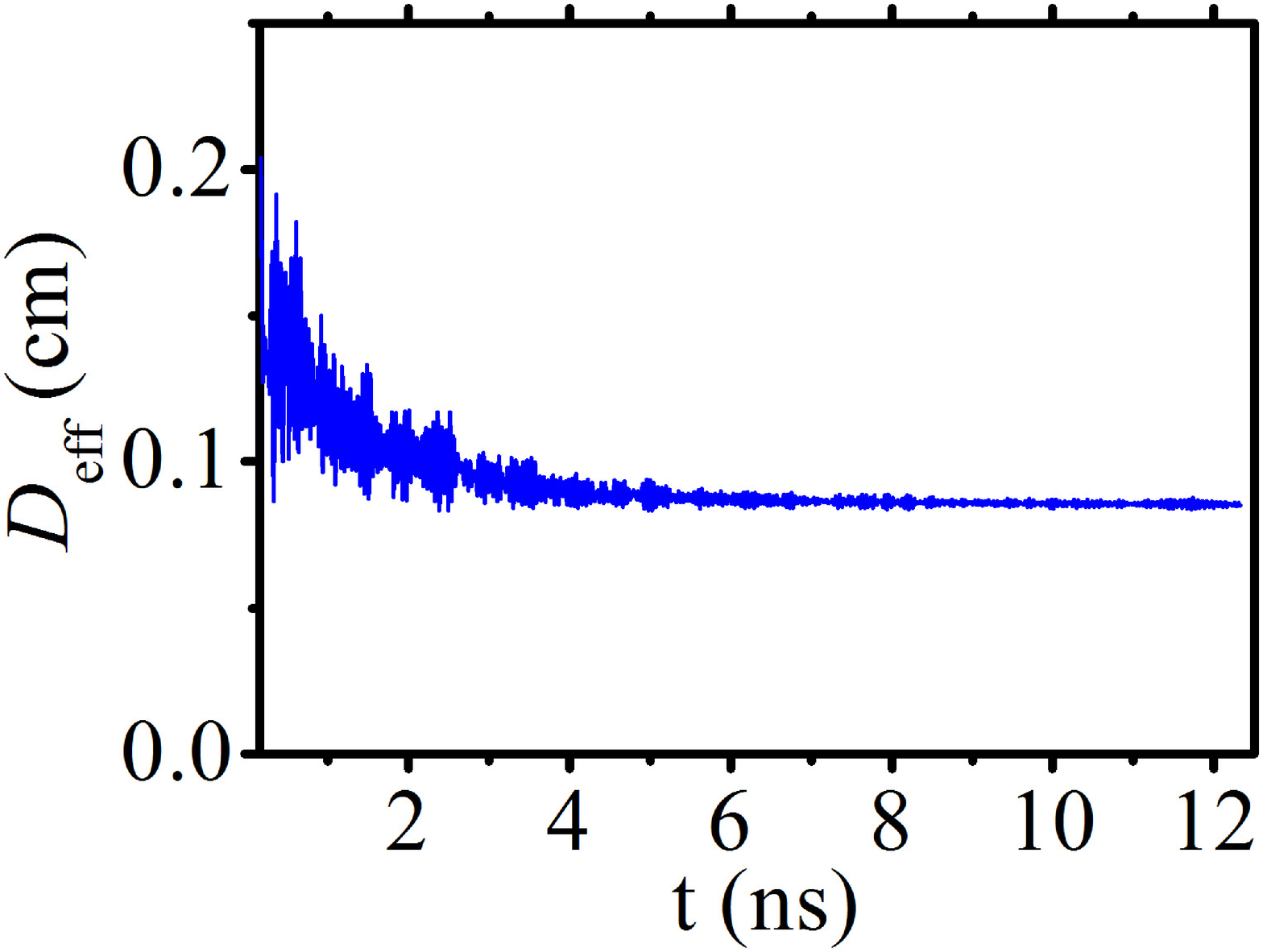}} %
\subfigure[]{\includegraphics[width=4cm, height=3cm]{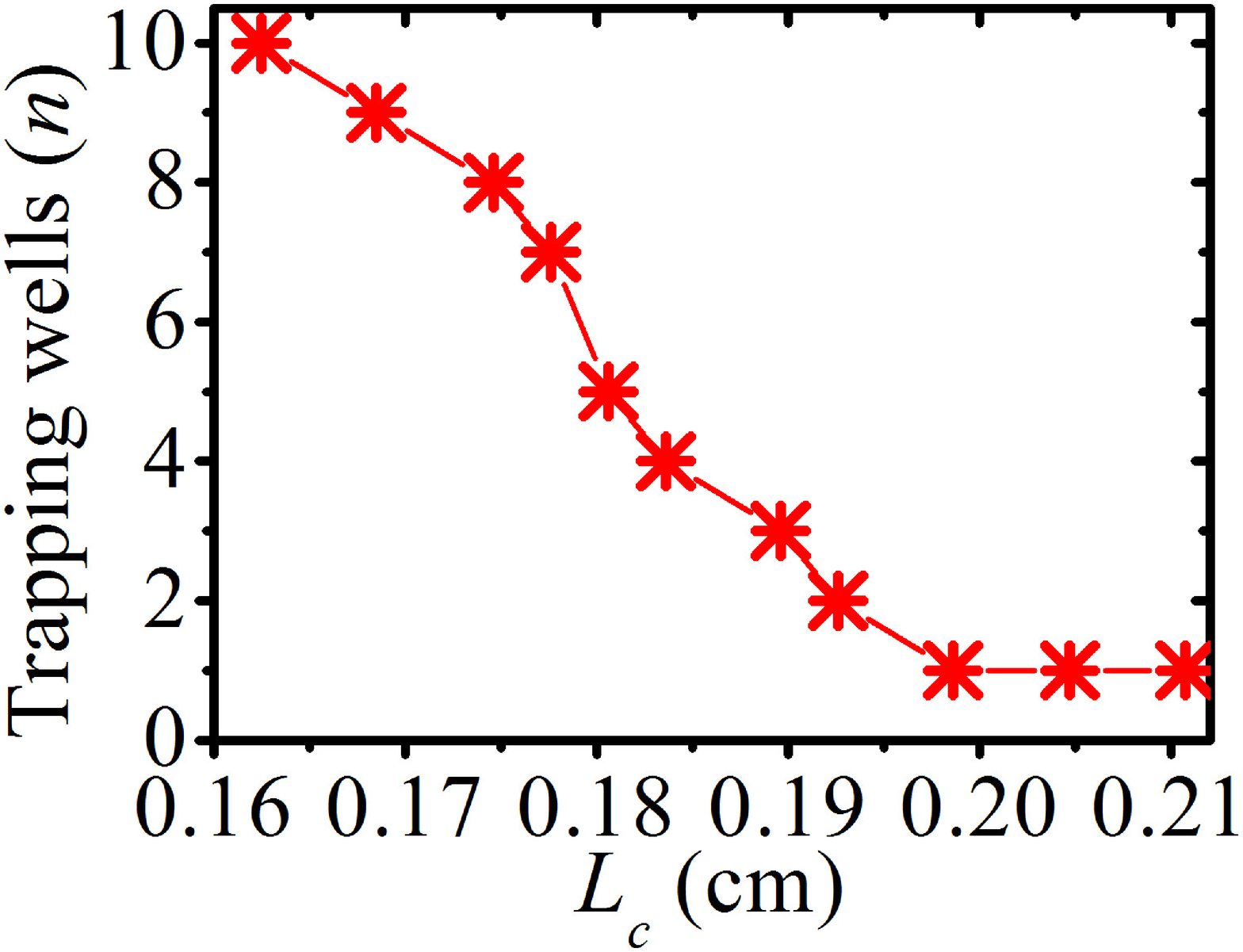}}
\caption{(Color online) (a) The temporal evolution of the effective squared
width of the pulse, defined as Eq. (\protect\ref{Deff}), with the same
parameters as in Figs. \protect\ref{fig2}(c,d). (b) The trapping site, $n$,
as a function of length $L_{c}$ of the chirped-BG segment, for the incident
intensity $I_{P}=2.72$ GW$/$cm$^{2}$ in the structure with $(S,d_{w},\protect%
\varepsilon )=(0.132$ cm, $50$ $\mathrm{\protect\mu }$m, $0.04)$.}
\label{fig4}
\end{figure}

The formation of the gap soliton from the initial pulse is additionally
illustrated in Fig. \ref{fig4}(a) by a typical example of the temporal
evolution of its effective squared spatial width, defined as
\begin{equation}
D_{\mathrm{eff}}(t)=\frac{\left[ \int_{0}^{L}\left(
|E_{f}(z,t)|^{2}+\left\vert E_{b}\left( z,t\right) \right\vert ^{2}\right) dz%
\right] ^{2}}{\int_{0}^{L}\left[ |E_{f}(z,t)|^{4}+|E_{b}(z,t)|^{4}\right] dz}%
{\LARGE .}  \label{Deff}
\end{equation}%
The width relaxes to a constant value by $t=4$ ns, when the pulse is already
trapped by a local potential well, and the soliton is well formed. Here, an
estimate for the distance travelled by the pulse before it forms a trapped
soliton is presented. Because $V_{sol}$ oscillates around zero before the
formation of a soliton completes [see the inset in Fig. \ref{fig2}(c)], we
assume that the average velocity of the pulse is $0.015c$, which corresponds
to an estimate for the travelled distance $\simeq 1.8$ cm, which is far
larger than the effective nonlinearity length, $L_{\mathrm{NL}}=1/\left(
\gamma I_{P}\right) \simeq 0.14$ cm.
Length $L_{c}$ of the chirped-BG segment strongly affects the subsequent
pulse trapping in the uniform-BG section. To illustrate this feature, a
relation between the trapping site, $n$, and $L_{c}$ is displayed in Fig. %
\ref{fig4}(b), for fixed parameters $(S,d_{w},\varepsilon )=(0.132$ cm$,50~%
\mathrm{\mu }$m$,0.04)$, and fixed incident intensity, $I_{P}=2.72$ GW$/$cm$%
^{2}$. In this case, the trapping occurs when $L_{c}$ takes values between $%
0.16$ cm and $0.21$ cm, with the trapping position, $n$, decreasing with the
increase of $L_{c}$ in this interval. This is explained by the fact that the
increase of $L_{c}$ leads to a reduction to the pulse's velocity, hence it
is captured earlier, at smaller $n$. However, the pulse cannot be trapped at
$L_{c}<0.16~$cm, and at $L_{c}>0.21$ cm. In the former case, the pulse's
velocity remains too large, allowing it to pass the entire uniform-BG
segment without being halted; in the latter case, the pulse cannot reach the
junction between the chirped-BG and uniform-BG segments, eventually
suffering decay in the chirped-BG segment, due to the strong BG-induced
dispersion.

\subsection{Multi-pulse trapping}

The present BG system is capable to trap several pulses too, in different
potential wells, provided that the pulses are launched into the system
successively. For instance, two-pulse trapping was demonstrated for the same
sets of parameters as considered above for the single-pulse case, i.e., $%
(S,d_{w},\varepsilon )=(0.132$ cm$,50$ $\mathrm{\mu }$m, $0.04)$.
Specifically, the first pulse with intensity $I_{P}=2.72$ GW$/$cm$^{2}$ [it
is expected to be trapped at the second potential well, according to Figs. %
\ref{fig2}(c,d)], is launched into the system at $t=0$, which is followed by
launching the second pulse with $I_{P}=2.07$ GW$/$cm$^{2}$ at $t=3$ ns
[Figs. \ref{fig2}(a,b) suggests that it may be captured by the first
potential well]. The outcome of the simulations demonstrates, in Figs.
5(a,b), that these two pulses are indeed stably trapped in two different
wells.

\begin{figure}[h]
\centering
\subfigure[]{\includegraphics[width=4cm, height=3cm]{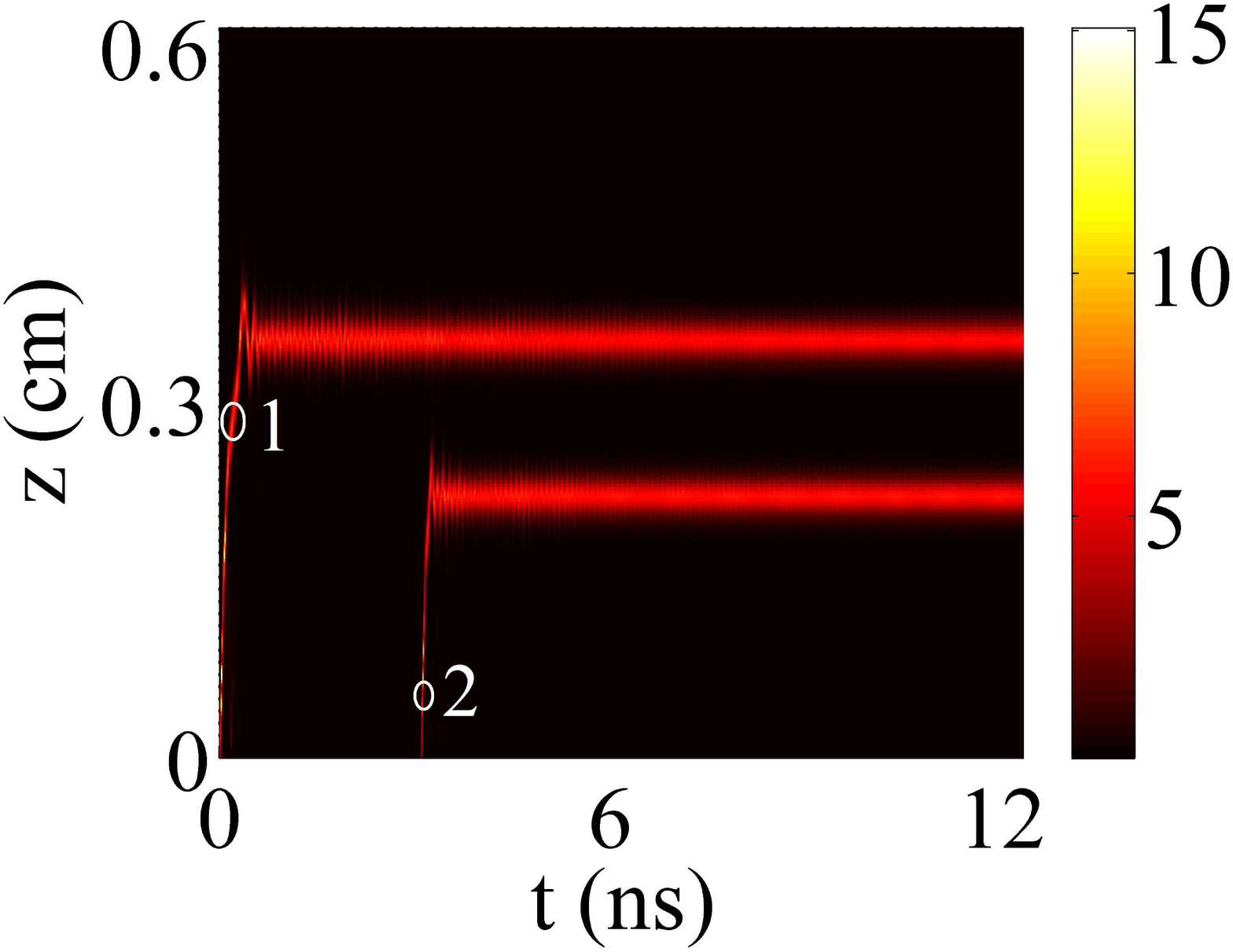}} %
\subfigure[]{\includegraphics[width=4cm, height=3cm]{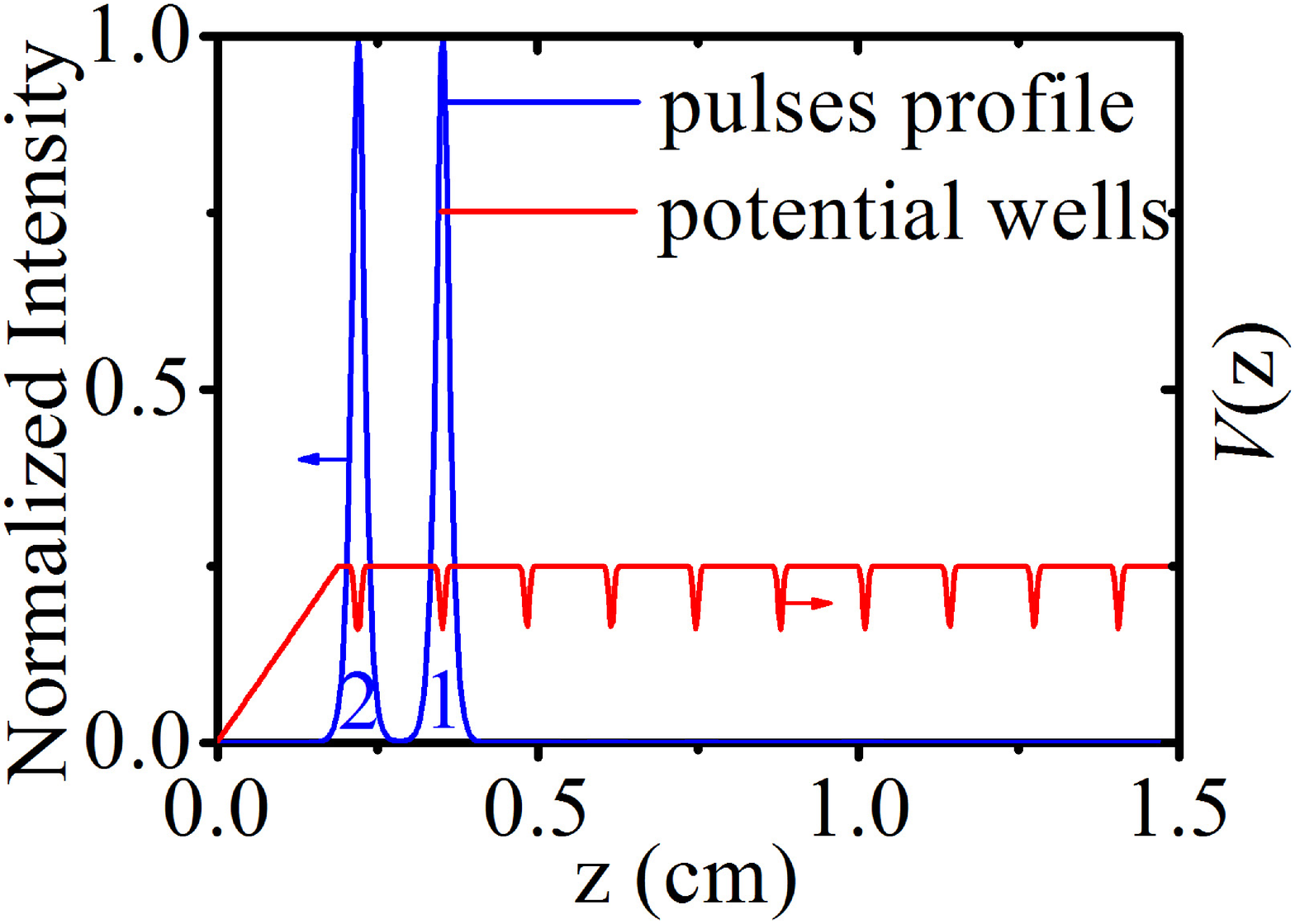}} %
\subfigure[]{\includegraphics[width=4cm, height=3cm]{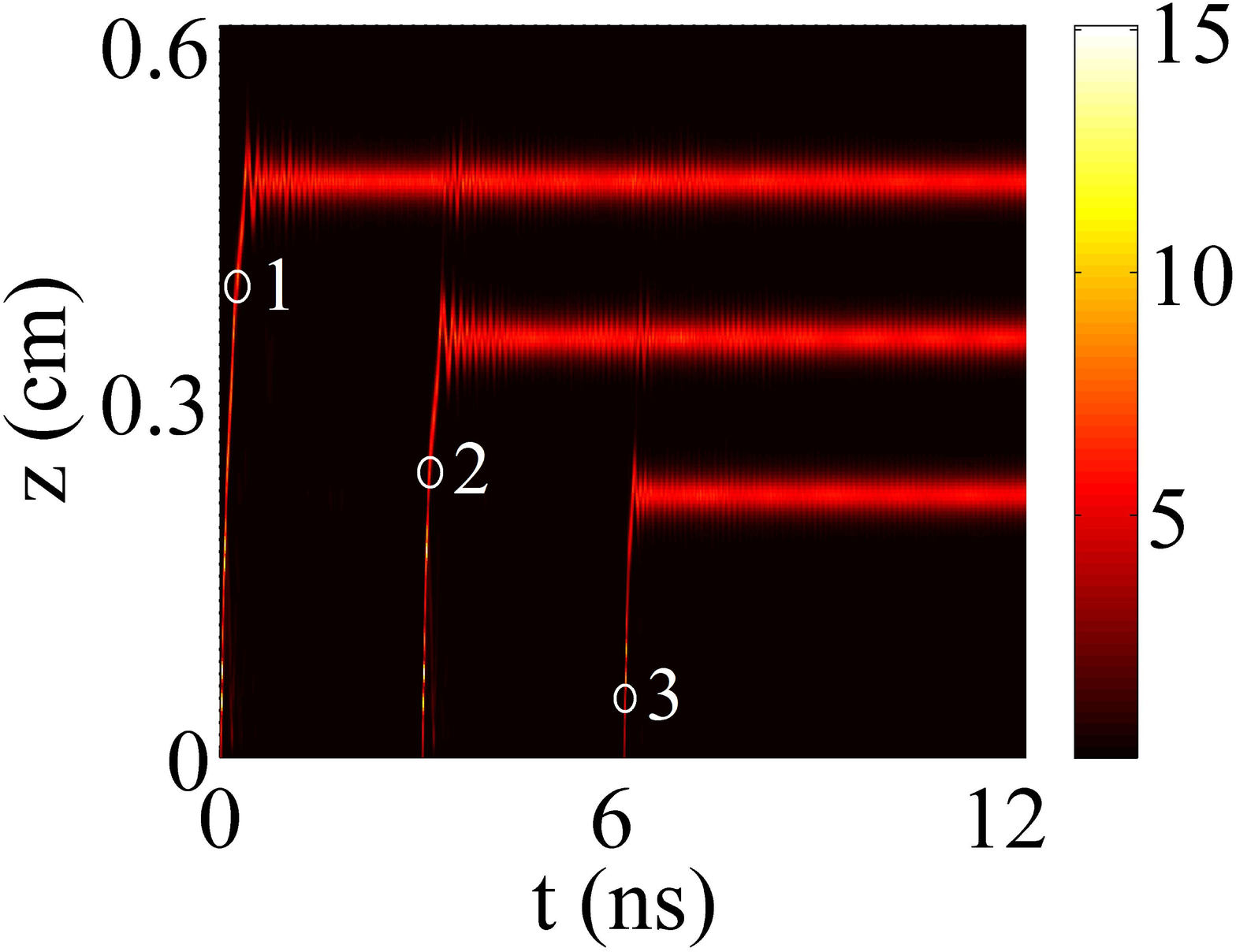}} %
\subfigure[]{\includegraphics[width=4cm, height=3cm]{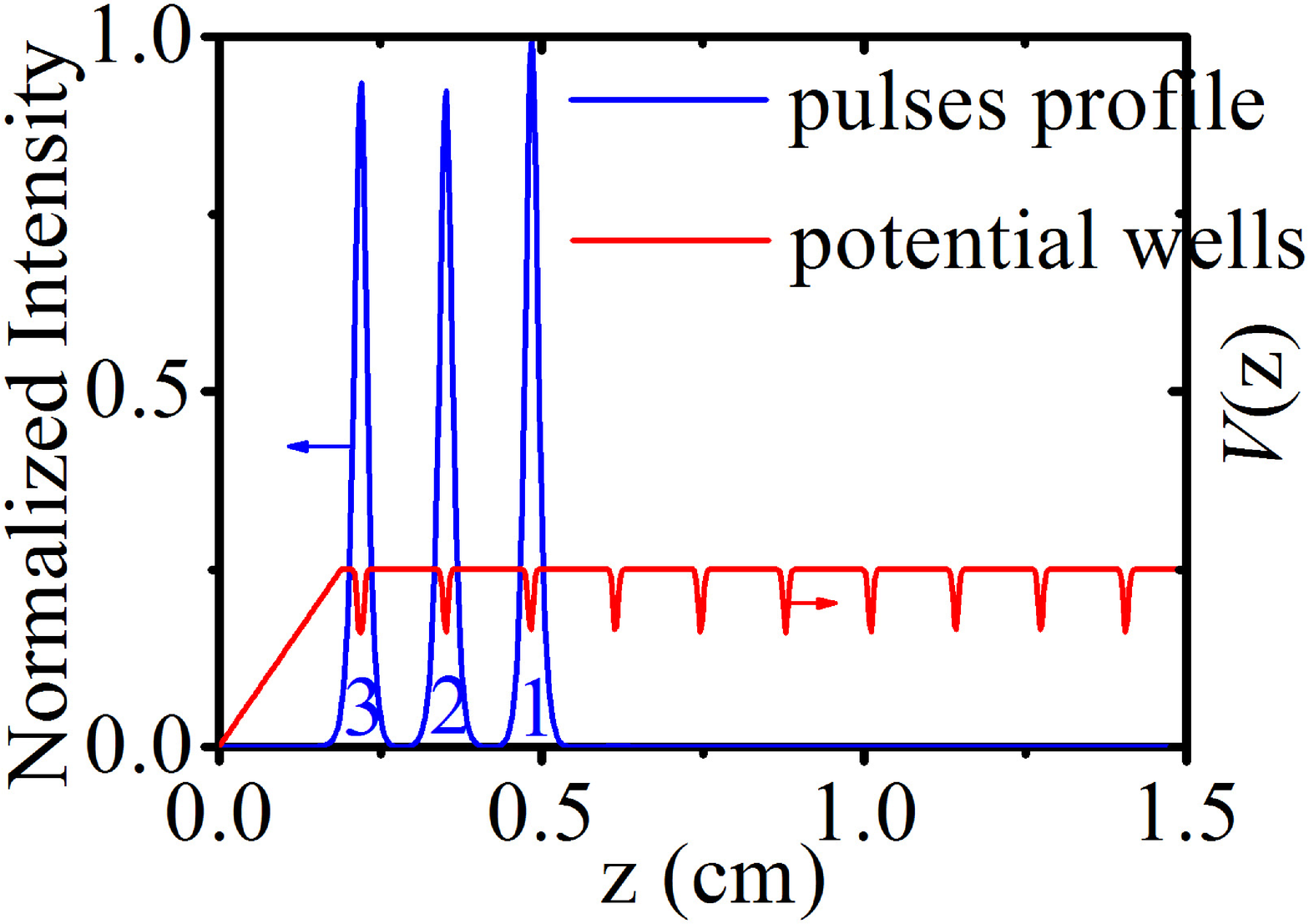}}
\caption{(Color online) Multi-pulse trapping in the Bragg-grating structure
with $(S,d_{w},\protect\varepsilon )=(0.132$ cm, $50$ $\mathrm{\protect\mu }$%
m, $0.04)$. (a) The trapping of two pulses with incident intensities of the first
and second pulses $I_{P}=2.72$ GW$/$cm$^{2}$ and $2.07$ GW$/$cm$%
^{2}$. The second pulse is launched into the gratings at $t=3$ ns. (c) The
trapping of three pulses, with intensities $I_{P}=3.04$ GW$/$cm$^{2}$, $2.78$
GW$/$cm$^{2}$, and $2.07$ GW$/$cm$^{2}$. The pulses are launched into the
gratings at $t=0$, $t=3$ ns, and $t=6$ ns, respectively. (b,d) Local-power
profiles for both sets at $t=12$ ns.}
\label{fig5}
\end{figure}

An example of three-pulse trapping is shown in Figs. \ref{fig5}(c, d). The
first pulse, with $I_{P}=3.04$ GW$/$cm$^{2}$, entering the chirped-BG input
edge at $t=0$, is trapped by the third wells. The second pulse with $%
I_{P}=2.78$ GW$/$cm$^{2}$ is launched at $t=3$ ns, and is captured by the
second well. The third pulse with $I_{P}=2.07$ GW$/$cm$^{2}$, launched at
$t=6$ ns, is eventually trapped in the first well, as suggested by the above
results for single pulses.

Although the above examples of the two- and three-pulse trapping seem as
simple superpositions of the single-pulse dynamical processes considered
above, it should be stressed that, loosing a considerable part of its power
on its way to the final halt [see Fig. \ref{fig3}(b)], each pulse leaves a
trace of emitted radiation in its wake. Nevertheless, Fig. \ref{fig5}
demonstrates that the passage of the areas ``contaminated" by the radiation
does not perturb the propagation of the secondary solitons, which
additionally attests to the solitons' stability.

\begin{figure}[t]
\centering
\includegraphics[width=8cm]{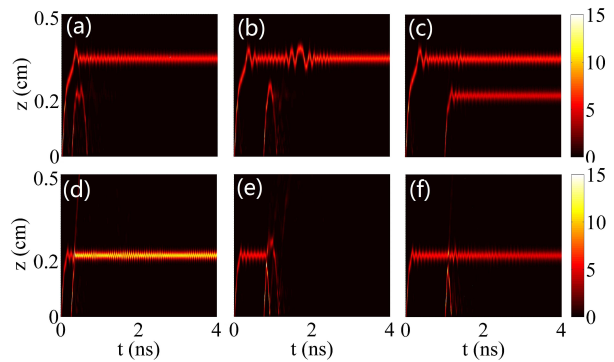}
\caption{(Color online) The two-pulse interaction in the system with $(S$,$%
d_{w},\protect\varepsilon )=(0.132$ cm, $50$ $\mathsf{\protect\mu }$m, $%
0.04) $. (a-c) The incident intensities of the first and second pulses $%
I_{P}=2.72$ GW/cm$^{2}$ and $2.07 $ GW/cm$^{2}$, while in (d-f) they are $%
I_{P}=2.07$ GW/cm$^{2}$ and $2.72$ GW/cm$^{2}$, respectively. The second
pulse is injected at $t=0.25$ ns (a,d), $0.75$ ns (b,e), and $1.0$ ns (c,f).}
\label{fig6}
\end{figure}

We also examined the case when the second pulse was launched right after the
first one, before the first pulse gets trapped, as in the example shown in
Figs. \ref{fig6}(a-c). In this case, it is concluded that, when the time
interval between these incident pulses is $<1$ ns, the second pulse bounces
back, see Figs. \ref{fig6}(a,b). On the other hand, if the time interval between the pulses exceeds
1 ns (allowing the first pulse to get well trapped), see Fig. \ref{fig6}(c)
[also Fig. \ref{fig5}(a)], these pulses can end up being stably trapped in
different potential wells simultaneously.
In addition, the situation where the incident intensities of the first and
second pulses are $I_{P}=2.07$ GW$/$cm$^{2}$ and $2.72$ GW$/$cm$^{2}$ is
demonstrated in Figs. \ref{fig6}(d-f). In this case, the two pulses collide
coherently in the course of the evolution.
Figure \ref{fig6}(d) demonstrates that these pulses interact attractively,
finally being trapped in the same potential well. On the contrary, Figs. %
\ref{fig6}(a,b,e,f), reveal repulsive interaction. Moreover, in Fig. %
\ref{fig6}(e), the repulsive interaction actually leads to release of an
initially trapped soliton, which may be used for retrieval of data bits
stored in the system, in terms of applications. The attractive and repulsive
sign of the interaction is well explained by the relative phase, namely,
they attract or repel each other when they are in phase or out of phase,
respectively \cite{RMP,Segev1999}.

\begin{figure}[t]
\centering
\includegraphics[width=8cm]{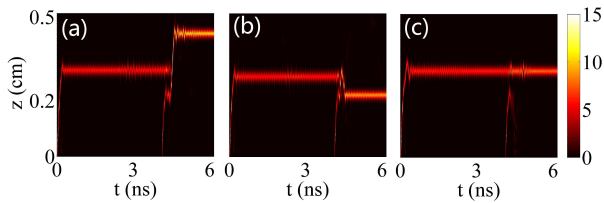}
\caption{(Color online) The two-pulse interaction in the system with the
reduced spacing of the lattice defect: $(S,d_{w},\protect\varepsilon %
)=(0.044$ cm, $50$ $\mathrm{\protect\mu }$m, $0.04)$ (a); $(0.066$ cm, $50$ $%
\mathrm{\protect\mu }$m, $0.04)$ (b); $(0.088$ cm, $50$ $\mathrm{\protect\mu
}$m, $0.04)$ (c). The first and second pulses are launched at $t=0$ and $%
t=4.1$ ns, with initial intensities $I_{P}=2.72$ GW$/$cm$^{2}$, and $%
I_{P}=2.07$ GW$/$cm$^{2}$, respectively.}
\label{fig7}
\end{figure}

It is possible to introduce interaction between solitons trapped by adjacent
defects, reducing the spacing, $S$, of the defect lattice [see Eq. (%
\ref{delta})]. Denser lattices are of obvious interest too in term of
applications. In the Figs. \ref{fig7}(a,b), and (c), reduced $S$ is taken as $%
S=0.044$ cm, $S=0.066$ cm and, additionally, $S=0.088$ cm, in comparison
with $S=0.132$ cm in Figs. \ref{fig5} and \ref{fig6}, while $d_{w}$ and $%
\varepsilon $ keep the same values as before. Although this reduction of the
spacing is not dramatic, it is sufficient to allow actual interactions between
solitons originally trapped in close potential wells, due to significant
overlapping between solitons' tails. For these settings, simulations were
again performed for the first pulse with $I_{P}=2.72$ GW$/$cm$^{2}$,
injected into the system at $t=0$, and the second pulse with $I_{P}=2.07$ GW$%
/$cm$^{2}$, injected at $t=4.1$ ns. The large temporal delay allowed the first pulse to settle
down into a trapped state before the second one would appear in the
vicinity. Figures \ref{fig7}(a) and (b) reveal that the interaction between
solitons, which are trapped, for a short time, by close defects, indeed
occurs in these cases, and changes the results.

In the case of $S=0.044$ cm, shown in Fig. \ref{fig7}(a), the second pulse
is originally trapped at distance $2S$ from the first one. In this case,
the attractively interacting pulses hop together into a higher potential
well and eventually merge in it. For $S=0.066$ cm, Fig. \ref{fig7}(b)
demonstrates that the first soliton, originally trapped in the second well,
is pulled by the incident second pulse back to the first well, where they
merge into a single pulse staying in the first well. On
the other hand, Fig. \ref{fig7}(c), corresponding to $S=0.088$ cm, shows that
the first pulse stays trapped in its original position, despite the repulsive
interaction with the second incident pulse. As mentioned above, the attractive
[Figs. \ref{fig7}(a,b)] or repulsive [Fig. \ref{fig7}(c)] sign of the interaction
is determined by the relative phase of the two pulses \cite{RMP,Segev1999}.

\section{Conclusions}

We have introduced a system engineered as a concatenation of linearly
chirped and uniform BGs (Bragg gratings), with the array of local defects
embedded into the uniform grating. The system of CMEs (coupled-mode
equations) have been used to simulate the evolution of single and multiple
pulses injected into the system. The systematic analysis has demonstrated
that, selecting parameters of the systems and the intensities of incident
pulse, the conversion of the pulse into a well-formed BG soliton and,
eventually, its trapping at a desired position, by one of the local defects,
are provided by the system. The relation of the trapping position and input
intensity was found, being close to a quadratic form. The co-trapping of two
or several pulses may be complicated by the strong interaction between them,
in the case when the temporal delay between the pulses, or the spacing
between adjacent local defects, is relatively small.

As a further development of the analysis, it may be interesting to study in
detail release of a trapped soliton by an incident one, an example of which
is displayed in Fig. \ref{fig6}(e). It may also be interesting to consider an
effects of a frequency shift added to the input pulse, which corresponds to
multiplying input (6) by $exp(-i\omega t)$, with constant frequency $\omega$.
On the other hand, for very slowsolitons, taking into account optoacoustic effects mediated by
electrostriction \cite{Tasgal1,Tasgal2} may improve the accuracy of the model.

\end{document}